\documentclass{aa501}
\usepackage{graphicx,natbib}
\bibpunct{(}{)}{,}{a}{}{,}
\begin{document}

   \title{Acoustic Oscillations in Solar and Stellar Flaring Loops}

   \author{V.M. Nakariakov \inst{1} \and D.
   Tsiklauri\inst{1}\fnmsep\thanks{Present address: Joule Physics Laboratory, School of Computing,
   Science and Engineering, University of Salford, Salford, M5 4WT,
   UK}
    \and A. Kelly \inst{1} \and
   T.D. Arber \inst{1} \and M.J. Aschwanden \inst{2}}
   \offprints{V.M. Nakariakov, \\ \email{valery@astro.warwick.ac.uk}}

   \institute{Physics Department, University of Warwick, Coventry,
   CV4 7AL, UK \and
Lockheed Martin, Advanced Technology
Center Solar \& Astrophysics Lab, Dept. L9-41, Bldg.~252, 3251
Hanover Street, Palo Alto, CA 94304, USA}

   \date{Received ???  2003}

\abstract{Evolution of a coronal loop in response to an impulsive
energy release is numerically modelled. It is shown that the loop
density evolution curves exhibit quasi-periodic perturbations with
the periods given approximately by the ratio of the loop length to
the average sound speed, associated with the second standing
harmonics of an acoustic wave. The density perturbations have a
maximum near the loop apex. The corresponding field-aligned flows
have a node near the apex. We suggest that the quasi-periodic
pulsations with periods in the range 10--300~s, frequently
observed in flaring coronal loops in the radio, visible light and
X-ray bands, may be produced by the second standing harmonic of
the acoustic mode.  }
\titlerunning{Acoustic Oscillations in Solar and Stellar Flaring Loops}
\authorrunning{Nakariakov et al.}
\maketitle

\section{Introduction}

Wave activity of the solar corona attracts attention in relation
with coronal heating and solar wind acceleration problems, and as
an efficient tool for MHD coronal seismology (e.g. Nakariakov
2003). The observational evidence of coronal waves and
oscillations is abundant. Low period coronal oscillations, in the
range between a few seconds to several minutes,  are believed to
be associated with magnetohydrodynamic waves. In particular,
propagating slow magnetoacoustic waves have been identified in
polar plumes (Ofman, Nakariakov \& DeForest 1999) and over loop
footpoints (Nakariakov et al. 2000, Tsiklauri \& Nakariakov 2001),
and standing global slow modes in loops (Ofman \& Wang 2002). The
majority of confidently interpreted examples of the coronal wave
activity has been found in the EUV coronal emission.

Observations in other spectral windows, in particular in the radio
band, also demonstrate various kinds of oscillations (e.g., the
quasi-periodic pulsations, or QPP, see Aschwanden 1987 for a
review), usually with periods from a few seconds to several tens
of seconds. It is commonly accepted that the waves with periods
about several seconds may be produced by either propagating or
standing sausage modes (Roberts, Edwin \& Benz 1984; Nakariakov,
Melnikov \& Reznikova 2003, Aschwanden, Nakariakov \& Melnikov
2004). However, it is known that the existence of the non-leaky
sausage mode cut-off imposes an upper limit on the possible wave
periods: waves with the periods longer than the cut-off period
cannot be trapped in the loops. For a magnetic cylinder of the
radius $a$, the cut-off period $P_\mathrm{c}$ is estimated (see,
e.g. Roberts et al. 1984) as
\begin{equation}\label{p_c}
  P_\mathrm{c} \approx \frac{2.62 a}{C_\mathrm{Ae}}
   \sqrt{\frac{C_\mathrm{Ae}^2-C_\mathrm{A0}^2}{C_\mathrm{s0}^2+ C_\mathrm{A0}^2}},
\end{equation}
where $C_\mathrm{s0}$ is the sound speed in the loop, and
$C_\mathrm{A0}$ and $C_\mathrm{Ae}$ are the Alfv\'en speeds inside
and outside the slab, respectively. For example, for typical
flaring loop parameters $a=6$~Mm, $C_\mathrm{A0}=1$~Mm/s,
$C_\mathrm{Ae} = 3C_\mathrm{A0}$ and $C_\mathrm{s0}=0.5$~Mm/s, the
cut-off period is about 15~s, and it is difficult to make it
greater than 20~s. However, X-ray band observations often give
much longer periodicities, frequently in association with a flare.
For example, periodicities from 20~s to 25~min have been presented
by Harrison (1987), McKenzie \& Mullan (1997) and Terekhov et al.
(2002). Similar periodicities have been observed in the decimeter
and microwave bands. In particular, Wang \& Xie (2000) observed
QPP with the periods of about 50~s at 1.42 and 2~GHz (in
association with an M4.4 X-ray flare). The coincidence of QPP
periods observed in the X-ray and in radio bands is not a
surprise, as the higher frequency radio bursts are found to
correlate very well with X-ray bursts (e.g., Benz \& Kane 1986).
Moreover, pulsations with the periods significantly greater than
the estimated sausage mode cut-off period have been found in both
hard X-ray and microwave bands simultaneously (e.g. Fu, Liu \& Li
1996, Tian, Gao \& Fu 1999).

Similar oscillations, with the period of 220~s, have recently been
observed in the white-light emission associated with stellar
flaring loops (Mathioudakis et al. 2003). Possibly, the 160~s
periodic oscillations observed by Houdebine et al. (1993) during a
flare on Ad-Leonis and 26 and 13~s coherent oscillations observed
by Zhilyaev et al. (2000) in EV Lac flares have the same nature.

A possible interpretation of these medium period QPPs may be
connected with kink or torsional modes (Zaitsev \& Stepanov 1989).
However, these modes are practically incompressible and, in the
case of a small amplitude, the produced perturbation of the
magnetic field is also very weak. (E.g., the direct observations
of kink modes in EUV, Aschwanden et al. 1999, Nakariakov et al.
1999). Thus, it is not simple to link these weak perturbations of
the magnetic field with observed QPPs.

In this study, we suggest an alternative mechanism for the
generation of long-period QPPs. We demonstrate that in a coronal
loop an impulsive energy release  generates the second spatial
harmonics of the acoustic mode. These oscillations are of high
quality and do not experience dissipation. This mode, producing
noticeable perturbations of the loop density and generating
field-aligned flows, is shown to be responsible for QPP with
medium and long periods.

\section{The model}

We describe plasma dynamics in a coronal loop by solving
numerically 1D radiative hydrodynamic equations (infinite magnetic
field approximation) that resemble closely  the Naval Research
Laboratory Solar Flux Tube Model (SOLFTM, Mariska 1987). The
numerical code that we use is a 1D version of Lagrangian Re-map
code (Arber et al. 2001) with the radiative loss limiters. As in
SOLFTM the coronal loop is connected with a dense, cold
(10$^4$~K), 5 Mm thick plasma region, that mimics the
chromosphere, and which because of its large density provides
sufficient amount of matter to fill the loop during the flare. The
model includes: the effects of gravitational stratification, heat
conduction, radiative losses, added external heat input, the
presence of Helium, non-linearity, and Braginskii bulk viscosity.
For the radiative loss function we use the form given in (Rosner
et al. 1978) extended to a wider temperature range (Peres et al.
1982, Priest 1982).

We performed a series of numerical runs simulating the response of
the loop to a flare-like impulsive heat deposition at a chosen
location. The heat deposition was modelled by the function
\begin{equation}\label{heat_func}
E_\mathrm{H}(s,t)=E_\mathrm{S}(s) E_\mathrm{T}(t)=
\end{equation}
$$
E_0 \exp\left(-{{(s-s_0)^2}\over{2 \sigma_\mathrm{s}^2}}\right)
\left[1+ Q_\mathrm{p} \exp\left(-{{(t-t_\mathrm{p})^2}\over{2
\sigma_\mathrm{t}^2}}\right) \right],
$$
where, $E_\mathrm{S}(s)$ and $E_\mathrm{T}(t)$ were spatial and
temporal parts of the heating function. $E_0$ is the amplitude in
units of erg cm$^{-3}$ s$^{-1}$; $s_0$ is the position of the
maximum heat deposition; $\sigma_\mathrm{s}$ is the heat
deposition localization scale. Function~(\ref{heat_func}) allows
us to model spatial and temporal profiles of the heat deposition
separately. As it is discussed below, the results are weakly
dependent upon the particular choice of this function. The only
restriction is the spatial and temporal characteristic scales to
be shorter than the loop length and the acoustic transit time
along the loop, respectively.

In the numerical runs presented here, the size of the heat
deposition region is $\sigma_\mathrm{s}=7$ Mm.  The flare peak
time is fixed in at 2200 s. The heat deposition duration,
$\sigma_\mathrm{t}$, is variable, from 42~s to 333~s. The heating
amplitude $Q_\mathrm{p}$ is fixed at $2 \times 10^4$, while $E_0$
is fixed at 0.004 erg cm$^{-3}$ s$^{-1}$. The loop length is taken
to be 55~Mm. We keep the coronal part of the loop initially at a
temperature of 1 MK and at a mass density of ${\rho} = \mu
m_\mathrm{p} n_\mathrm{e} = 6.6 \times 10^{-16}$ g cm$^{-3}$ (at
the loop apex), for a helium-to-hydrogen number density ratio of
0.05, i.e., with the mean molecular weight $\mu=1.1$. This
corresponds to an electron density of $n_\mathrm{e}=3.6\times
10^8$ cm$^{-3}$. In our approach, specific initial conditions in
region connecting corona to the chromosphere and chromosphere
itself have little or no effect on the corona dynamics as they are
rapidly modified, self-consistently, according to the radiative
hydrodynamics equations. Resolution in all numerical runs was
fixed at 1000 grid points, which were distributed non-uniformly in
order to properly resolve strong gradients in the transition
region. Even tripling the number of grid points does not alter the
numerical results, thus assuring full convergence of the
simulation.

\section{Acoustic second standing harmonics}

\begin{figure}
\centering
  \resizebox{\hsize}{!}{\includegraphics{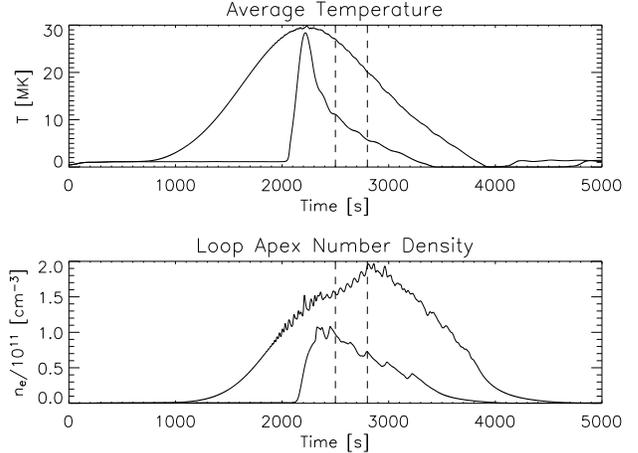}}
\caption{Time evolution of the average temperature (upper panel)
and the number density at apex (lower panel) in the flaring loop
in the case of apex heating. The upper curves correspond to longer
flares and the lower to the shorter flares. The vertical straight
lines show the time interval of particularly distinct oscillations
on the upper curve.} \label{f1}
\end{figure}

The typical loop evolution is shown in Figure~\ref{f1}. The
impulsively heated loop responds with an increase of the
temperature. The particular response is determined by the duration
of the heat deposition $\sigma_\mathrm{t}$. In Fig.~\ref{f1} the
cases of $\sigma_\mathrm{t}=42$~s and 333~s are presented. The
chromospheric evaporation increases loop density with up-flows of
the order of a hundred km s$^{-1}$. Then, during the peak of the
flare, the combined action of the heat input and the conductive
and radiate losses, yields an oscillatory flow pattern with
typical amplitudes of up to few tens of km s$^{-1}$. In the
cooling phase, the matter drains out of the loop exhibiting
oscillations; and the loop finally reaches an equilibrium.

In the context of this study, the most interesting feature in
Fig.~1 is the clear quasi-periodic oscillations of the density
time-profiles, especially about and after the peak of the flare.
This behaviour closely resembles the radio and X-ray flaring
light-curves (see the citations in Introduction).

Typical time-distance plots of the evolution of the density and
the velocity are shown in Figure~2. Here, the slow, in comparison
with the oscillation period, aperiodic variations of the density
and the velocity are subtracted. The velocity perturbation clearly
demonstrates the node near the apex, while the density
perturbation is maximal there (anti-node). This behaviour is
consistent with the second standing acoustic mode,

 \begin{eqnarray}\label{ther}
 V_x(s,t)=A \cos \left( {{2 \pi C_\mathrm{s}}\over{L}} t\right) \sin
 \left(
 {{2 \pi}\over{L}} s\right),\\
  \rho(s,t)=-{{A \rho_0}\over{C_\mathrm{s}}} \sin
 \left( {{2 \pi C_\mathrm{s}}\over{L}} t\right) \cos \left(
 {{2 \pi}\over{L}} s\right),
 \end{eqnarray}
where $C_\mathrm{s}$ is a speed of sound, $A$ is wave amplitude,
$L$ is loop length, and $s$ is a distance along the loop with the
zero at the loop top.

\begin{figure}[]
  \resizebox{\hsize}{!}{\includegraphics{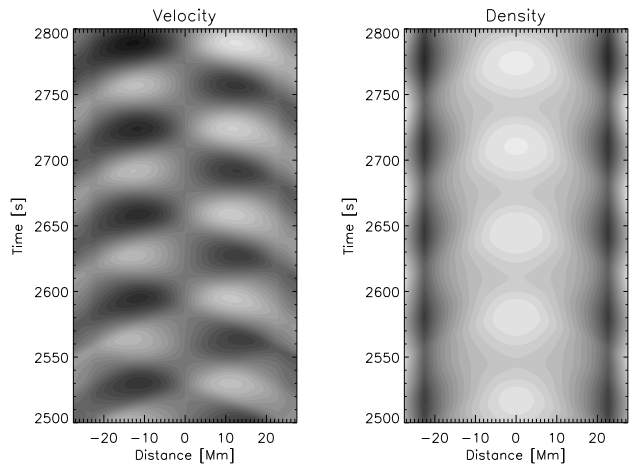}}
\caption{The time-distance plots of perturbations of the velocity
and density, after subtraction of the slow evolution of the
background, for the time interval highlighted for the upper curve
in Fig~1.} \label{f2}
\end{figure}

\begin{figure}[]
\resizebox{\hsize}{!}{\includegraphics{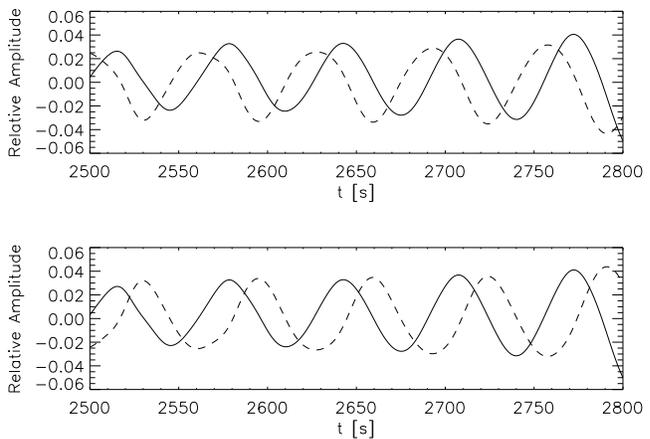}}
\caption{Time signatures of the velocity and density perturbations
outputted at the points $-6$ Mm (top panel) and $+6$ Mm (bottom
panel) of the loop length for the same time interval as
Fig~\ref{f2}. The solid curve shows the plasma number density in
units of $10^{11}$ cm$^{-3}$. The dashed line gives the velocity
normalized to 400 km s$^{-1}$. Here, the slow variations of the
density and the velocity are subtracted.} \label{f3}
\end{figure}

The top panel of Figure~\ref{f3} shows characteristic time
signatures of the density and velocity oscillations at two points
symmetric with respect to the loop apex. Depending upon the leg of
the loop, the phase shift between the density and the velocity
oscillations is plus or minus quarter period, as it is expected
for the second standing acoustic harmonic (see Eqs.~(\ref{ther})).
Perturbations of the density are in phase at both  points of
measurement. The oscillations are quasi-harmonic, with a
well-defined period. In the discussed case, the period of the
oscillation is 64 s. According to Eqs.~(\ref{ther}), the period of
the second spatial harmonic should be $P=L/C_\mathrm{s}$. The
practical formula for the determination of the oscillation period
is
\begin{equation}\label{2sah}
  P/\mathrm{s} \approx 6.7 \times (L/\mathrm{Mm})/\sqrt{(T/\mathrm{MK})}
\end{equation}
where $T$ is the average temperature in the loop. (This expression
is a particular case of the general formula suggested by Roberts
et al. (1984), who pointed out that the second standing harmonic
could likely be excited in coronal loops.) Substituting the loop
length $L=55$~Mm (see Fig.~2) and the average temperature, 25~MK
(see top left panel in Fig.~1 in the range of 2500--2800~s), we
obtain 73.7~s, which is reasonably close to the result of the
simulations. The  discrepancy of about 15\% is explained by the
fact that the actual reflection points are situated over the loop
footpoints (see Fig.~\ref{f2}) because of stratification. This
reduces  slightly the effective length of the loop and,
consequently the resonant period of the mode. However, as the
uncertainty in the observational determination of the loop length
is comparable with this error, formula~(\ref{2sah}) provides a
satisfactory estimation of the oscillation period.

\section{Conclusions}

We suggest that the second standing acoustic mode may be
responsible for QPP with  periods of about 10--300~s (estimated by
Eq.~(\ref{ther})) observed in solar and stellar flare light
curves. This mechanism is similar to the interpretation of coronal
loop oscillations observed with SUMER, proposed by Ofman \& Wang
(2002). The main new element in our study is connected with the
mode excitation. We demonstrate that the second standing acoustic
harmonic appears as \textit{a natural response} of the loop to an
impulsive energy deposition. The SUMER oscillations are likely to
be associated with some other excitation mechanism, as only a
small fraction of SUMER oscillations are observed in association
with solar flares (Wang et al. 2003).

Traditionally, the acoustic wave interpretation was excluded as
these waves were supposed to be highly dissipative. However,
numerical simulations discussed above, as well as recently gained
abundant observational evidence of the presence of acoustic waves
in the solar corona, suggest that the observed periodicities can
be associated with this mode. The physical mechanism responsible
for the induction of the quasi-periodic pulsations can be
understood in terms of auto-oscillations generated by an
electric-circuit generator. Indeed, the physical system modelled
here contains all the necessary ingredients of a generator: the DC
power supply (thermal instability), the nonlinear element (the
plasma) and the resonator (the loop). This may explain why the
oscillations may be observed to be dissipationless. However,
proper analytical theory of the excitation of this mode is still
to be developed.

The oscillation period is determined by the ratio of the loop
length and the average sound speed in the loop. The typical
amplitude of the density and velocity perturbations is 2-10\% of
the background. This value is consistent with observed amplitudes
of X-ray QPPs (McKenzie \& Mullan 1997). However, the observed
amplitudes of radioband QPPs are sometimes higher. This
discrepancy would be resolved by taking into account the specific
mechanism responsible for the modulation of radio emission, and
the line-of-sight angle (e.g. Cooper, Nakariakov \& Williams
2003). In certain condition, the modulation mechanism can amplify
the pulsations up to the observable level.

The model developed here is quite approximate, as it does not take
into account two-dimensional MHD effects such as centrifugal force
and the perturbation of the loop cross-section. However, we
believe that the neglected effects do not change the qualitative
picture described here and that Eq.~(\ref{2sah}) gives a correct
estimation for the oscillation period.

Finally, we would like to emphasize that the generation of the
second spatial harmonic of an acoustic wave, by means of which we
try to explain the observed quasi-periodic oscillations in flaring
loops, is a consistent feature, seen for a wide range of physical
parameters, including the case of non-symmetric heating functions.
More detailed analysis will be presented elsewhere.

\begin{acknowledgements}
Numerical calculations of this work were performed using the PPARC
funded Compaq MHD Cluster at St Andrews and Astro-Sun cluster at
Warwick. This work was supported in part by PPARC (DT) and EPSRC
(AK). The authors are grateful to the referee, Mihalis
Mathioudakis, for valuable comments.
\end{acknowledgements}

\end{document}